\documentclass[final]{aipproc}
\usepackage{makeidx}
\usepackage{epsfig}
\usepackage{graphicx}

\input{aipcheck}
\layoutstyle{8x11single}
\graphicspath{{./fig2/}}

\begin{document}

\title{Theory of stochastic transitions in area preserving maps}

\author{Massimo
Tessarotto\thanks{email: M.Tessarotto@cmfd.univ.trieste.it}$\ \
^{a,c} $ and Piero Nicolini\thanks{email:
Piero.Nicolini@cmfd.univ.trieste.it}$\ \ ^{a,b}\ $} {address={\
$^{a} $Department of Mathematics and Informatics,University of Trieste, Italy\\
$^{b} $National Institute of Nuclear Physics (INFN) Trieste
Section, Italy\\
 $^{c} $Consortium for Magnetofluid Dynamics\thanks{Web site: http://cmfd.univ.trieste.it}, University
of Trieste, Italy}}

\begin{abstract}
A famous aspect of discrete dynamical systems defined by
area-preserving maps is the physical interpretation of stochastic
transitions occurring locally which manifest themselves through
the destruction of invariant KAM curves and the local or global
onset of chaos. Despite numerous previous investigations (see in
particular Chirikov \cite{Chirikov1979}, Greene \cite{Greene79},
Percival \cite {Percival79}, Escande and Doveil \cite{Escande81}
and MacKay \cite{MacKay}) based on different approaches, several
aspects of the phenomenon still escape a complete understanding
and a rigorous description.  In particular Greene's approach is
based on several conjectures, one of which is that the stochastic
transition leading to the destruction of the last KAM curve in the
standard map is due the linear destabilization of the elliptic
points belonging to a peculiar family of invariants sets $\left\{
I(m,n)\right\} $ (rational iterates) having rational winding
numbers and associated to the last KAM curve.  Purpose of this
work is to analyze the nonlinear phenomena leading to the
stochastic transition in the standard map and their effect on the
destabilization of the invariant sets associated to the KAM
curves, leading, ultimately, to the destruction of the KAM curves
themselves.
\end{abstract}

\maketitle

%$^{\star }$} \url{http://cmfd.univ.trieste.it}.

%%%%%%%%%%%%%%%%%%%%%%%%%%%%%%%%%%%%%%%%%%%%
%% MAINMATTER
%%%%%%%%%%%%%%%%%%%%%%%%%%%%%%%%%%%%%%%%%%%%

\section{Introduction: phenomenology of stochastic transitions}

The phenomenology of stochastic transitions for canonical (and
area-preserving) maps, such as the standard map
\cite{Chirikov1979}, is one of the most popular subjects of
investigation in nonlinear dynamics. In particular, in the case of
the transition to global chaos this amount to the search of the
the so-called \emph{last KAM curve, } i.e. a continuous invariant
curve which separates nearby subdomains of the phase-space
($\Gamma$) characterized at most only by local chaos. The search
of such a KAM curve and the conditions of its possible destruction
for arbitrary canonical maps may result potentially a difficult
problem. In the past, sufficient conditions for the global
stochastic transition, for the standard map and suitable
families of area-preserving maps, have been proposed, such as Chirikov Criterion \cite%
{Chirikov1979}) or Greene's Chaotic Hypothesis \cite{Greene79}.
However, despite various attempts (see in particular Chirikov\cite{Chirikov1979},
Greene \cite{Greene79}, Percival %
 \cite{Percival79}, Escande and Doveil \cite{Escande81} and MacKay \cite{MacKay}%
) based on different approaches, several aspects of the phenomenology of
stochastic transitions still escape a complete understanding and a rigorous
description.

In two dimensions closed KAM curves{\ can be closed (so-called invariant
circles) or open. In the first case they separate phase space }since
trajectories cannot cross KAM curves, and hence produce the phenomenon of
phase-space confinement in phase space. Also open KAM curves, however, can
separate space. Discrete invariant sets $J$ {\ do not separate space and
therefore cannot give rise to phase-space confinement. }In general one
expects that the winding number of an arbitrary closed (or open) KAM curve,
in particular those which separate space, depend on the stochasticity
parameter $K,${\ \ i.e., }
\[
{\alpha =\alpha (J,K).}
\]

Therefore, a KAM curve can be destroyed for appropriate values of
$K,$ i.e., when $\alpha (J,K)$ becomes rational. This is called
local stochastic transition of the KAM curve $J.$ The local
destruction of a KAM surface, and consequent disappearance of
local phase-space confinement, is also called transition to local
chaos. In particular, there may exist a special KAM curve, the
so-called \emph{last KAM curve}, whose destruction permits the
existence of phase-space trajectories which ''connect'' all
boundaries lines of phase space $\Gamma ,$ {\ thus destroying
''global'' phase-space confinement. The stochastic transition of
this KAM curve is called as global stochastic transition or
transition to global chaos.

\section{A classification of area-preserving maps}

For the purpose of attempting a systematic investigation of
area-preserving maps, generated by discrete dynamical systems, a
convenient approach is to construct them, in analogy to continuous
dynamical systems, in terms of a variational principle. Indeed the
same concept of Hamiltonian system can be extended to discrete
dynamical systems in such a way to yield, by construction,
area-preserving (i.e., conservative) maps. By analogy to
continuous theory, discrete Hamiltonian systems can be defined by
means of a suitable discrete variational principle expressed by a
variational equation of the form $\delta S(\mathbf{x)=}0$ and
denoted as ''\textit{discrete modified Hamilton variational
principle}''$.$ As usual, this equation must
be satisfied by arbitrary syncronous variations of the form $\delta \mathbf{%
x}_{i}=\left( \delta \mathbf{q}_{i},\delta \mathbf{p}_{i}\right) $ (with $%
i=0,n),$ being $\delta \mathbf{x}_{i}\equiv \mathbf{x}_{i}-\mathbf{x}_{1i}$
and $\mathbf{x}_{i},\mathbf{x}_{1i}$ two arbitrary elements of the
syncronous functional class
\begin{equation}
\left\{ \mathbf{x}\right\} =\left\{ \mathbf{x}_{i}:\mathbf{x}_{i},%
i\in N,\mathbf{x}_{o}=\mathbf{x}^{A},\mathbf{x}_{n}=\mathbf{x}^{B},\mathbf{x%
}_{i}=(\mathbf{q}_{i}\mathbf{,p}_{i}\mathbf{)\in }\Gamma \mathbf{\subseteq }%
R^{2g}\right\} .
\end{equation}%
where $\mathbf{x}_{o}=\mathbf{x}^{A},\mathbf{x}_{n}=\mathbf{x}^{B}$ are
suitable compatible boundary conditions and $\mathbf{q=(}q_{1},..,q_{g}%
\mathbf{),p=(}p_{1},..,p_{g}\mathbf{),}$ being $g$ an integer $g\geq 1$ to
be identified with the degree of freedom$\mathbf{.}$ A convenient definition
of the variational functional is
\begin{equation}
S(\mathbf{x)=}\sum_{i\in 0,n}\left\{ \left( \mathbf{q}_{i+1}-\mathbf{q}%
_{i}\right) \cdot \mathbf{p}_{i+1}-H(\mathbf{q}_{i},\mathbf{p}%
_{i+1})\right\} ,  \label{funzHam.discreto}
\end{equation}%
where $H(\mathbf{q}_{i},\mathbf{p}_{i+1})$ denotes a suitably smooth real
function, here denotes as \textit{discrete Hamiltonian function}. The
corresponding Eulero-Langrange equations read therefore:
\begin{equation}
\delta S(\mathbf{x)=}\sum_{i\in 0,n}\delta \mathbf{p}_{i+1}\cdot \left[
\left( \mathbf{q}_{i+1}-\mathbf{q}_{i}\right) -\frac{\partial }{\partial
\mathbf{p}_{i+1}}H(\mathbf{q}_{i},\mathbf{p}_{i+1})\right] +
\end{equation}%
\[
+\sum_{i\in 0,n}\delta \mathbf{q}_{i}\cdot \left[ \left( \mathbf{p}_{i}-%
\mathbf{p}_{i+1}\right) -\frac{\partial }{\partial \mathbf{q}_{i}}H(\mathbf{q%
}_{i},\mathbf{p}_{i+1})\right] =0,
\]%
where the partial derivatives should be meant in the sense
\begin{equation}
\frac{\partial }{\partial \mathbf{p}_{i+1}}H(\mathbf{q}_{i},\mathbf{p}%
_{i+1})=\left. \frac{\partial }{\partial \mathbf{p}}H(\mathbf{q}_{i},\mathbf{%
p})\right| _{\mathbf{p}=\mathbf{p}_{i+1}},
\end{equation}%
\begin{equation}
\frac{\partial }{\partial \mathbf{q}_{i}}H(\mathbf{q}_{i},\mathbf{p}%
_{i+1})=\left. \frac{\partial }{\partial \mathbf{q}}H(\mathbf{q},\mathbf{p}%
_{i+1})\right| _{\mathbf{q}=\mathbf{q}_{i}}.
\end{equation}%
Hence, they deliver the \textit{discrete canonical map} $T$
\begin{equation}
\left\{
\begin{array}{c}
\mathbf{q}_{i+1}=\mathbf{q}_{i}+\frac{\partial }{\partial \mathbf{p}_{i+1}}H(%
\mathbf{q}_{i},\mathbf{p}_{i+1}), \\
\mathbf{p}_{i+1}=\mathbf{p}_{i}-\frac{\partial }{\partial \mathbf{q}_{i}}H(%
\mathbf{q}_{i},\mathbf{p}_{i+1}), \\
\mathbf{x}_{o}=\mathbf{x}^{A},\mathbf{x}_{n}=\mathbf{x}^{B},
\end{array}%
\right.   \label{eq.Hamilton discrete}
\end{equation}%
(defined for $i=0,n$) which can be proven to satisfy the fundamental Poisson
brackets:%
\begin{equation}
\left[ \mathbf{q}_{i+1},\mathbf{p}_{i+1}\right] _{(\mathbf{q}_{i}\mathbf{,p}%
_{i})}=\underline{\underline{\mathbf{1}}},
\end{equation}%
\begin{equation}
\left[ \mathbf{q}_{i+1},\mathbf{q}_{i+1}\right] _{(\mathbf{q}_{i}\mathbf{,p}%
_{i})}=\left[ \mathbf{p}_{i+1},\mathbf{p}_{i+1}\right] _{(\mathbf{q}_{i}%
\mathbf{,p}_{i})}=\underline{\underline{\mathbf{0}}}.
\end{equation}%
This implies, b{y virtue of the previous definition (\ref{funzHam.discreto})
for the variational functional, that discrete Hamiltonian systems, just like}
continuous ones, result trivially area--preserving. In particular, this
property is satisfied by arbitrary elements of the sequence $\left\{
x_{i},i\in N\right\} _{\mathbf{x}^{A}},$ defining the {phase-space
trajectory of the discrete canonical system,\ which is determined by the map
}$T$ ,{\ for an arbitrary initial state }$\mathbf{x}_{o}=\mathbf{x}^{A}\in
\Gamma $ (and letting $\mathbf{x}_{n}=\mathbf{x}^{B}).${\ }

Conversely, in the particular case of two-dimensional (i.e., for $g=1),$ it
follows that area-preserving maps $T$ of the form
\begin{equation}
{\ }\mathbf{x}_{i}^{{\ }}{\equiv (q}_{i}{,p}_{i}{\ )\rightarrow \mathbf{x}%
_{i+1}^{{\ }}{\equiv }(q}_{i+1}{,p}_{i+1}{)\equiv T}\mathbf{x}_{i}{,}
\end{equation}
are necessarily canonical and hence of the type defined above (\ref%
{eq.Hamilton discrete}).

\subsection{Parameter-dependent canonical maps}

In the sequel we shall consider, without loss of generality,
parameter-dependent 2D canonical maps $T${\ characterized by a bounded phase
space $\Gamma \equiv $\textit{\ }$[-\pi $,$\pi \lbrack \times \lbrack 0,2\pi
\lbrack $ and having a discrete Hamiltonian function of the form }$%
H=H(q_{i},p_{i+1},K),$ being $K$ a real parameter (to be denoted as \textit{%
stochasticity parameter}) and $H$ a real analytic function of $K.$ Moreover,
we shall impose that the discrete Hamiltonian $H$ satisfies the constraint
\begin{equation}
{\ }H(q_{i},p_{i+1},K=0)=H_{0}(p_{i+1}{),}
\end{equation}%
for arbitrary  $i=0,n$ and phase-space trajectories, which implies
\begin{equation}
H(q_{i},p_{i+1},K)=H_{0}(p_{i+1}{)+}\sum_{j=1}^{\infty
}K^{j}H_{j}(q_{i},p_{i+1}).
\end{equation}%
An example of 2D maps of this type, which results also a
\textit{twist map,}
i.e., satisfies the twist condition%
\begin{equation}
\frac{\partial q_{i+1}}{\partial p_{i+1}}\neq 0,
\end{equation}
is provided by the so-called standard map [\cite{Chirikov1979},\cite%
{Greene79}].
\begin{eqnarray}
\ q_{i+1} &=&\ q_{i}+p_{i+1} \\
{\ }p_{i+1} &=&\ p_{i}+K\sin q_{i}{.}  \nonumber
\end{eqnarray}%
Since the map is manifestly area-preserving, it results also canonical, the
discrete Hamiltonian function being in this case:
\begin{equation}
H(q_{i},p_{i+1},K)=\frac{{1}}{{2}}p_{i+1}^{2}+K\cos q_{i}.
\end{equation}
Two-dimensional parameter-dependent area-preserving maps of this type are
characterized by discrete and continuos invariant sets (the latter called
KAM curves). \ In particular, for $K=0$ all continuous subsets of $\Gamma $
defined by the equations
\begin{equation}
p_{i}=p^{A}
\end{equation}%
are necessarily invariant (KAM curves). For $K>0$ KAM curves can
change shape or disappear altogether.

In particular, discrete invariant sets may appear which are
characterized generally by a finite number ($n $ denoted as order
of the set) of {\ }${m}${-periodic invariant points (i.e., points
which are returned to themselves after }$m$ iterations of the
map). Discrete and continuous
invariant sets can be distinguished by means of their winding numbers ${%
\alpha (\mathbf{x}^{A})}$, i.e.: \

\begin{equation}
{\ \alpha (\mathbf{x}^{A})}=\lim_{{\ k\rightarrow \infty }}\frac{{\ \delta }%
^{(k)}{\ (\mathbf{x}^{A}\ )}}{{\ 2\pi k}}
\end{equation}
{\ where }$\delta ^{(k)}({\alpha (\mathbf{x}^{A})}$ {\ is the rotation angle
after at the }$k-${\ th iteration of the map }$T,$ namely$:$
\begin{equation}
{\ \delta }^{(k)}{(\mathbf{x}^{A})=}\sum_{{\ j=1,k-1}}\left( {\ p}_{j+1}{\ -p%
}_{j}\right)
\end{equation}
{\ and} $x_{j}$\ {\ is the }$j-${\ th image of
}${\mathbf{x}^{A}}${\ through the canonical map }$T.$ As a
consequence, it is immediate to prove that discrete invariants
sets of order $n$ and periodicity $m,$ to be denote by the symbol
$I(m,n)\equiv \left\{ {\ }m,n,K,{\mathbf{x}^{A}}\right\} ,$
display necessarily a rational winding number ${\alpha
(\mathbf{x}^{A})=m/n,} $ whereas KAM\ curves have necessarily
irrational winding numbers. In both cases, one expects generally\
that winding numbers result function of the stochastic parameter
$K$, i.e. that $\alpha =\alpha(J,K)$, where $J$ denotes an
arbitrary invariant set. In particular one may expect that a given
KAM curve $\gamma $ which exists for $K=0,$ and is characterized
by an irrational winding number, may become a discrete invariant
set in correspondence of a suitable critical values of the
stochasticity parameter $K$ the function $\alpha (J,K)$ may become
rational. This is called local stochastic transition. A basic
conjecture (Greene, 1979 \cite{Greene79}) is that for arbitrary
area-preserving maps each KAM curve $\gamma \ ${\ can actually be
constructed as a limit set of an appropriate family of
}${m}${-periodic discrete invariant sets }$\left\{
I(m,n),m=m(n),n\in N\right\} ,$ {\ i.e.,
for }$n\rightarrow \infty :$%
\begin{equation}
{\ I(m(n),n)\rightarrow \gamma ,}
\end{equation}
{\ \ where the discrete invariant sets }$I(m(n),n)$ {\ are
characterized by
the distinctive property that their winding numbers }$\alpha (I(m,n))=\frac{%
m(n)}{n}${\ \ tend to }$\alpha (\gamma )$ {\ in the asymptotic limit, i.e.,}
\begin{equation}
\lim_{{\ n\rightarrow \infty }}{\ \alpha (I(m,n))}=\lim_{{\ n\rightarrow
\infty }}\frac{{\ m(n)}}{{\ n}}={\ \alpha (\gamma ).\ }
\end{equation}

A basic conjecture proposed by Greene \cite{Greene79}) is that for
arbitrary area-preserving maps each KAM curve $\gamma ,$ {\ having winding number }$%
\alpha (\gamma ),\ ${\ can actually be constructed as a limit set
of an appropriate family of m-periodic discrete invariant sets (to
be denoted as {\it rational iterates}) }$\left\{
I(m,n),m=m(n),n\in N\right\} ,$ {\ i.e., for }$n\rightarrow \infty :$%
\begin{equation}
{\ I(m(n),n)\rightarrow \gamma ,}
\end{equation}
{\ \ where the discrete invariant sets }$J(m(n),n)$ {\ are
characterized by
the distinctive property that their winding numbers }$\alpha (I(m,n))=\frac{%
m(n)}{n}${\ \ tend to }$\alpha (\gamma )$ {\ in the asymptotic
limit, i.e.,}
\begin{equation}
\lim_{{\ n\rightarrow \infty }}{\ \alpha (I(m,n))}=\lim_{{\
n\rightarrow \infty }}\frac{{\ m(n)}}{{\ n}}={\ \alpha (\gamma ).\
}
\end{equation}

\begin{figure}[tbp]
\includegraphics[height=.3\textheight]{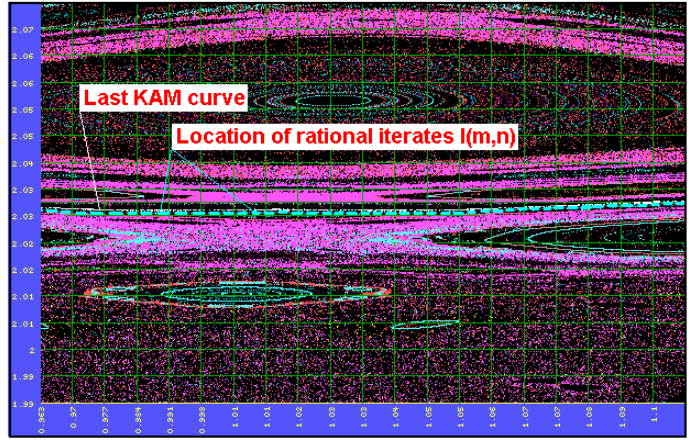}
\caption{Location of last KAM curve and of the rational iterates}
\end{figure}
\section{Greene's chaotic hypothesis}

The method developed by Greene \cite{Greene79} for investigating stochastic
transitions in area-preserving maps (and in particular for the standard map)
is based on the following set of conjectures, denoted as \textsl{Chaotic
Hypothesis}:\newline

{\bf A)} \textsl{1st conjecture (irrational winding number of the
last KAM curve)}: the last KAM surface is characterized by the
irrational winding number
\begin{equation}
\alpha (\gamma )=\alpha _{GM}\equiv \frac{\sqrt{5}-1}{2};
\end{equation}
(}$\alpha _{GM}\equiv ${\ golden mean) which can be represented by the
continuous fraction}
\begin{equation}
\frac{\sqrt{5}-1}{2}=\frac{1}{1+\frac{1}{1+\frac{1}{1+..}}}.
\end{equation}

{\bf B)} \textsl{2nd conjecture (mechanism of destruction of a KAM
curve) :} the disappearance of a KAM surface (in particular the
''last KAM surface'' whose destruction produces the global
stochastic transition) is related to the linear destabilization of
the invariant points (elliptic equilibrium points) of an
appropriate family of m-periodic discrete invariant sets $I(m,n),$
{\ i.e. when the Lyapunov exponents of the invariant points become
positive for suitable values of the stochasticity parameter }$K${\
. The stochastic transition can be estimated by analyzing the
bifurcation conditions (linear destabilization) of the invariant
sets }$I(m,n)$ {\ for sufficiently high values of the order }$n${\
\ of these sets. In particular, for sufficiently high values of
}$n${\ \ and for all }$n_{1}\geq n$ {\ all points of the sets}
$I(m(n_{1}),n_{1})$ {\ become linearly unstable.\emph{\ }}

{\bf C)} \textsl{3rd conjecture (invariant sets): }in the case of
the last KAM surface of the standard map the discrete invariant
sets $J(m,n)$ {\ \ coincide with the so called rational
iterates}$.$ {\ The winding numbers of these special invariant
sets are given by the following sequence, obtained truncating the
previous continuous fraction, i.e.,}

%\QTP{
%}
\begin{equation}
\alpha (I(1,2))=\frac{1}{1+1}=\frac{1}{2},...,\alpha (I(m,n))=\frac{m(n)}{n}%
,\alpha (I(n,n+m))=\frac{n}{n+m},...\rightarrow \alpha _{GM},
\end{equation}
{\ \ yielding the family of sets}
\begin{equation}
\left\{ {\ I(m(n),n),n\in N}\right\} \equiv \left\{ {\
I(1,2),I(2,3),I(3,5),...,I(m(n),n),...}\right\}
\end{equation}
{\ where }$n,m$ {\ are Fibonacci numbers belonging to the sequence}
\begin{equation}
{\ n,m\in }\left\{ {\ 2,3,5,8,13,...k}\right\} ,
\end{equation}
{\ with }$k${\ \ given by the sum of the previous two elements of the
sequence. Here m and n denote, respectively, the periodicity and the order
(number of points) of the generic set of the family }$I(m,n).$

Using numerical experiments Greene was able to justify at least in part
there conjectures and achieve the following results for the standard map:

{\ I) \textsl{identification of the last KAM curve and the family
of rational iterates: (see figure 2)}}$\left\{ I(m,n),n\in
N\right\} $ {\ (see figure 5): near stochastic transition the
invariant sets }$I(m,n)$ {\ are found to be localized below the
last KAM; these sets continue to exist also after transition
(i.e., for values of the stochasticity parameter }$K$ { \ larger
than the critical value }$K_{crit}$ {\ ).}

{\ II) \textsl{linear stability analysis of rational iterates
}}$I(m,n)${\ \textsl{: highest order }}$n$ {\ \textsl{considered
for the rational iterate }}$n=233,${\ \textsl{\ i.e.,
}}$I(m,n)=I(144,233)$

{\ III) \textsl{estimate of the critical value of the
stochasticity parameter }}$K$ {\ \textsl{at stochastic
transition:}}

\begin{equation}
{\ K\cong 0.9716,}
\end{equation}
{\ determined by the requirement that the elliptic invariant points of the
set }$I(144,233)$ {\ become linearly unstable}.

Although these results have been confirmed by other authors using
different approaches (Percival\cite{Percival79}, Escande and
Doveil\cite{Escande81} and MacKay\cite{MacKay}), they leave
unexplained several aspects and basic issues of the theory. In
particular, the identification of the invariant sets $I(m,n)$ {\
(conjecture B) associated to the last KAM curve remains not
obvious even for the standard map.} {\ \ In the same way this
problem remains unanswered for arbitrary KAM curves and more
general area-preserving maps.

\section{The Nonlinear Chaotic Hypothesis (NCH)}

In this regard, a natural question is whether the family of
invariant sets $I(m,n)$ (the rational iterates), associated to a
given KAM curve, i.e., for which the KAM curve is a limit set, is
unique or not. This issue arises immediately when trying to relate
Greene's chaotic hypothesis to the well known Chirikov criterion
for the overlapping of resonances (Chirikov Criterion
\cite{Chirikov1979}). In fact, the latter states that KAM curves
are destroyed when the distance between the two lowest order
resonances which encompass the same KAM curve, i.e., lie on
opposite sides of the KAM curve, becomes comparable to the sum of
the semi-amplitudes of the separatrix islands formed by the two
resonances. On the other hand the existence of multiple sequences
of discrete invariants sets having the same limit set (a given KAM
curve) does not contradict the original Greene conjecture, which
does not specify the precise location of the invariant sets
$J(m,n)${\ nor exclude their possible non-uniqueness. This has
motivated us to search for possible candidates for these invariant
sets, associated in principle to an arbitrary KAM curve, which are
denoted here as {\it alternate rational iterates} $Y(m_{j},n)$}}.
Numerical experiments performed for the stochastic transition of
the last KAM curve of the standard map \cite{Tessarotto2004}
confirm this conjecture and suggest a new formulation of the
Greene's Chaotic Hypothesis, replacing conjectures B and C with
the following three new conjectures B1 and C1:
\newline {\bf B1)} \textsl{new 2nd conjecture (invariant
sets related to a KAM curve) :} the disappearance of a KAM surface
$\gamma $ {\ (in particular the ''last KAM surface'' which
determines with its destruction the global stochastic transition)
is related to the existence to two families of discrete invariant
sets }$I(m,n)$ {\ and }$Y(m_{j},n)${\ , with }$j ${\ \ a suitable
integer}, {\ defined in such a way their winding numbers tend to
}$\alpha (\gamma ),${\ \ i.e.,}
\begin{equation}
\lim_{{\ n\rightarrow \infty }}{\ \alpha (I(m,n))}=\lim_{{\ n\rightarrow
\infty }}\frac{{\ m(n)}}{{\ n}}={\ \alpha (\gamma ),\ }
\end{equation}
\begin{equation}
\lim_{{\ n\rightarrow \infty }}{\ \alpha (Y(m}_{j}{\ ,n))}=\lim_{{\
n\rightarrow \infty }}\frac{{\ m}_{k}{\ (n)}}{{\ n}}={\ \alpha (\gamma ).\ }
\end{equation}

{\bf C1)} \textsl{new 3rd conjecture (invariant sets related to
the last KAM curve of the standard map) : }in the case of the last
KAM surface of the standard map the discrete invariant sets
$I(m,n)$ {\ and }$Y(m_{j},n)${\ , \ coincide, respectively, with
the sets of rational and alternate rational iterates when letting
}$j=1.$\newline

{\bf D1)} \textsl{new 4th conjecture (nonlinear mechanism of
destruction of a KAM curve): }the stochastic transition of the
last KAM surface occurs when
the Euclidean distance between corresponding elliptic points of the sets $%
I(m,n)$ {\ and }$Y(m_{j},n),$ {\ considered as a function of} {\ the
stochasticity} $K,$ {\ becomes minima, for sufficiently high values of }$n$.%
\newline

The set of hypotheses A (Greene's first conjecture), together with
B1,C1 and D1, is here denoted here as Nonlinear Chaotic Hypothesis
(NCH). We stress that the conjecture D1 \textsl{(new 4th
conjecture) appears consistent with (but differs from) Chirikov
Criterion, since the distance between nearby resonances cannot
exceed the sum of the semi-amplitudes of the separatrix islands
formed by the two resonances. Moreover, it replaces the assumption
on linear stability with a non-linear destabilization criterion,
i.e. the requirement that or the Euclidean distance between
adjacent elliptic points belonging respectively to two sets
}$I(m,n)$ and $Y(m_{1},n)$ having the same order $n$. In fact, in
principle it is not obvious at all why a linear stability
criterion (as Greene's 4th conjecture) should hold in general
since an elliptic invariant point might result at the same time
linearly unstably and nonlinearly stable. In addition, invariant
points of different invariants sets (for example adjacent points
invariants sets of the same order $n$, $I(m,n)$ {\ and
}$Y(m_{1},n)$), may not become all linearly unstable
simultaneously, i.e. for the same value of the stochasticity
parameter $K$. \\
Therefore, NCH should permit, in particular, not only a more
accurate evaluation of the critical value of the stochasticity
parameter $K$ occurring at stochastic transition, but also a
precise identification of the families of discrete invariant sets
associated, in principle, to an arbitrary KAM curve.  Finally, the
new approach closes the gap between Greene's Chaotic Hypothesis
and Chirikov criterion of overlapping of resonances, permitting at
the same time a deeper understanding of the mechanism of
stochastic transitions in area-preserving maps.

%%%%%%%%%%%%%%%%%%%%%%%%%%%%%%%%%%%%%%%%%%%%%%%%
%% BACKMATTER
%%%%%%%%%%%%%%%%%%%%%%%%%%%%%%%%%%%%%%%%%%%%%%%%

\begin{theacknowledgments}
Work developed in the framework of the PRIN Research Program
``Programma Cofin 2002: Metodi matematici delle teorie
cinetiche''( MIUR Italian Ministry) and partially supported (for
P.N.) by the National Group of Mathematical Physics of INdAM
(Istituto Nazionale di Alta Matematica), (P.N) by the INFN
(Istituto Nazionale di Fisica Nucleare), Trieste (Italy) and
(M.T.) by the Consortium for Magnetofluid Dynamics, University of
Trieste, Italy.
\end{theacknowledgments}

%\QTP{
%}
%%%%%%%%%%%%%%%%%%%%%%%%%%%%%%%%%%%%%%%%%%%%%%%%
%% You may have to change the BibTeX style below, depending on your
%% setup or preferences.
%%
%% If the bibliography is produced without BibTeX comment out the
%% following lines and see the aipguide.pdf for further information.
%%
%% For The AIP proceedings layouts use either
%%%%%%%%%%%%%%%%%%%%%%%%%%%%%%%%%%%%%%%%%%%%

%\begin{references}

% if natbib is available
%\bibliographystyle{aipprocl} % if natbib is missing

%%%%%%%%%%%%%%%%%%%%%%%%%%%%%%%%%%%%%%%%%%%
%% You probably want to use your own bibtex database here
%%%%%%%%%%%%%%%%%%%%%%%%%%%%%%%%%%%%%%%%%%%
%\bibliographystyle{aipproc}
%\bibliography{sample}

\begin{thebibliography}{8}
\bibitem{Chirikov1979}  { B.V.Chirikov, Phys. Reports 52, 262 (1979). }

\bibitem{Greene79}  { J. Greene, J.Math.Physics{\bf \ 20, }1183 (1979)%
{\bf .} }

\bibitem{Percival79}  { I.C. Percival, in {\it Nonlinear Dynamics and
the beam-beam interaction,} by M.Month and J.C.Herrera Eds.,
AIP Conference Proceedings No.67, American Institute of
Physics, New York, 1979) }

\bibitem{Escande81}  { D.C.Escande and F.Doveil, Phys.Lett {\bf 83A},
307 (1981), Phys.Lett. {\bf 84A}, 399 (1981) and
J.Stat.Physics{\bf \ 26}, 257 (1981). }

\bibitem{MacKay}  { R.S.MacKay, {\it Renormalization in area
preserving maps}, Ph.D. Thesis Princeton 1982, World
Scientific (1993). }

\bibitem{Tessarotto2004}  { M.Tessarotto and P.Nicolini, to be
submitted (2004).}
%\end{references}
\end{thebibliography}

%%%%%%%%%%%%%%%%%%%%%%%%%%%%%%%%%%%%%%%%%%%
%% Just a reminder that you may have to run bibtex
%% All of it up to \end{document} can be removed
%% if you don't like the warning.
%%%%%%%%%%%%%%%%%%%%%%%%%%%%%%%%%%%%%%%%%%%
%\IfFileExists{\jobname.bbl}{} {\typeout{} \typeout{******************************************}
%\typeout{** Please run
%"bibtex \jobname" to optain}
%\typeout{** the bibliography and then re-run
%LaTeX} \typeout{** twice to fix the references!}
%\typeout{******************************************} \typeout{} }

\end{document}